\documentclass{svproc}
\usepackage{url}

\usepackage{graphicx}
\usepackage{multirow}
\usepackage{booktabs}
\usepackage{caption}
\usepackage{subcaption}
\usepackage{times}

\begin{document}
\mainmatter              % start of a contribution

\title{Glioblastoma Multiforme Patient Survival Prediction}
\titlerunning{Glioblastoma Multiforme}  % abbreviated title (for running head)

\author{Snehal Rajput\inst{1} \and Rupal Agravat\inst{2} \and Mohendra Roy\inst{1}
 \and Mehul S. Raval\inst{2}}

\authorrunning{Snehal Rajput et al.} % abbreviated author list (for running head)

\tocauthor{Snehal Rajput, Rupal Agravat, Mohendra Roy, Mehul S. Raval}
\institute{Department of Information and Communication Technology, \\ Pandit Deendayal Petroleum University, Gandhinagar, India. \\ \and
School of Engineering and Applied Science, \\ Ahmedabad University, Ahmedabad, India. \\ \email{[snehal.rphd19,mohendra.roy]@sot.pdpu.ac.in} \\ \email{mehul.raval@ahduni.edu.in}, \email{rupal.agravat@iet.ahduni.edu.in}}

\maketitle              % typeset the title of the contribution

\begin{abstract}
\vspace{-5mm}

Glioblastoma Multiforme is a very aggressive type of brain tumor. Due to spatial and temporal intra-tissue inhomogeneity, location and the extent of the cancer tissue, it is difficult to detect and dissect the tumor regions. In this paper, we propose survival prognosis models using four regressors operating on handcrafted image-based and radiomics features. We hypothesize that the radiomics shape features have the highest correlation with survival prediction. The proposed approaches were assessed on the Brain Tumor Segmentation (BraTS-2020) challenge dataset. The highest accuracy of image features with random forest regressor approach was 51.5\% for the training and 51.7\% for the validation dataset. The gradient boosting regressor with shape features gave an accuracy of 91.5\% and 62.1\% on training and validation datasets respectively. It is better than the BraTS 2020 survival prediction challenge winners on the training and validation datasets. Our work shows that handcrafted features exhibit a strong correlation with survival prediction. The consensus based regressor with gradient boosting and radiomics shape features is the best combination for survival prediction.

% We would like to encourage you to list your keywords within
% the abstract section using the \keywords{...} command.
\end{abstract} 
\footnotetext[1]{All authors have contributed equally to this work.}

\section{Introduction}
Glioblastoma multiforme (GBM) is the commonest type of primary malignant brain tumor. In the case of adults, glioblastoma makes up 60\% of all brain tumors \cite{taylor2019glioblastoma}. The World Health Organization (WHO) classified GBM as a grade IV type of cancer due to its invasive and diffusive nature. Patients suffering from GBM have a poor prognosis, with a median survival rate of about ten months \cite{taylor2019glioblastoma}. This is due to its aggressive nature, highly heterogeneous appearance, location, shape, and unpredictable response to therapy \cite{ronneberger2015u}.  %GBM has  complex clinical behavior, and it is difficult to predict the disease outcome. 

Magnetic Resonance Imaging (MRI) has been widely utilized to examine tumors due to its non-hazardousness, high contrast and superior resolution. Generally, manual segmentation of a tumor in MRI is time consuming and prone to subjective error. In this regards an automated segmentation method would be of enormous help to oncologists and clinicians. It can help in early diagnosis as well as in therapeutic strategy planning. In recent years, deep learning-based segmentation approaches have outperformed traditional state-of-the-art methods \cite{zhao2019bag,mckinley2019triplanar}. Segmentation delineates the brain tumor into Whole Tumor (WT), Enhancing Tumor (ET), and Tumor Core (TC). Handcrafted features extracted from these segments are used to classify the survival days of the patients. %into long-term-survivors, mid-term-survivors, and short-term-survivors. 

% \begin{figure}[!ht]
% \vspace{-5mm}
% \centering
% \begin{subfigure}{.30\textwidth}
% \includegraphics[width=3cm, height=3cm]{fig/axial.png}
% \caption{}
% \end{subfigure} \hfill
% \centering
% \begin{subfigure}{.30\textwidth}
% \includegraphics[width=3cm, height=3cm]{fig/coronal.png}
% \caption{}
% \end{subfigure} \hfill
% \centering
% \begin{subfigure}{.30\textwidth}
% \includegraphics[width=3cm, height=3cm]{fig/sagittal.png}
% \caption{}
% \end{subfigure} 
% \caption{3D MRI views: (a) Axial (b) Coronal (c) Sagittal}
% \label{fig:fig1}
% \vspace{-6mm}
% \end{figure}

There are many segmentation models available. Recently, Jiang et al. \cite{jiang2019two}, in the BraTS 2019 challenge, proposed a two-stage asymmetry cascaded U-Net \cite{ronneberger2015u} structure. Each model is made up of a larger encoder  in order to be able to extract more complex semantic features and a smaller decoder part for generating a segmentation map with a size identical to the input. Zhao et al. \cite{zhao2019bag} proposed multiple methods to generate robust segmentation results. They grouped it into data processing, model devising, and optimization modules. Multiple methods are assimilated into each of these modules to enhance segmentation results. McKinley et al. \cite{mckinley2019triplanar} proposed a Densenet based U-Net architecture. Convolutions that were dilated were used to bring about an increase in the receptive field, which retains spatial information. The model was trained by combining label uncertainty loss, binary cross-entropy and focal loss. Dice scores on the BraTS-2019 validation dataset were 0.91(WT), 0.83(TC), 0.77(ET), and on the BraTS-2019 test dataset were 0.89(WT), 0.83(TC), 0.81(ET). Therefore, researchers seem to be favouring the U-Net based architecture for segmentation.

Once the tumor is segmented, features are extracted for overall survival prediction. Agravat et al. \cite{agravat2019brain} used dense layers U-Net trained on the focal loss for segmentation. Next, age, statistical features and radiomic features train the Random Forest Regressor (RFR) for survival prediction and the obtained accuracy on the test dataset was 0.58. Wang et al. \cite{wang2019automatic} used U-Net and U-Net ensembles with attention gates trained on soft dice scores and cross-entropy segmentation. For survival prediction, they proposed the following prognosis models: i) baseline model where only the age feature was used to train a linear regressor model. ii) Radiomic model where morphological and texture features were extracted from segmentation results. iii) Tumor invasiveness model, where relative invasiveness coefficient (RIC) and age feature train the support vector regressor model. The tumor invasive model was found best for survival prediction. The accuracy for survival prediction was 0.59 and 0.56 for BraTS-2019 validation and test dataset respectively. Feng et al. \cite{feng2019brain} used an ensemble of U-Net models. The models were trained on patches having brain pixels. The main advantage of using an ensemble method is that the network parameter need not be fine-tuned. Further, for OS prediction, volume and surface area features were extracted for each Region of Interest (ROIs) and age to train a linear regression model. The training and testing set accuracy was reported as 0.31 and 0.55 respectively on the BraTS-2019 datasets. Wang et al. \cite{wang20193d} utilized a 3D U-Net-based model, and the training occurred in two phases using patching methods. The first phase included both brain and background pixels, whereas the second included only brain pixels. The dice score coefficient loss function was utilized to train the 3D U-Net model. Further for survival prediction, volume, surface area and age were used to train the ANN model. The training, validation, and testing accuracy of the models were 0.515, 0.448, and 0.551 respectively. Islam et al. \cite{islam2019brain} proposed a 3D U-Net architecture for segmentation, where attention blocks have been desegregated with the decoder modules. For survival prediction, various geometric, fractal, and histogram-based features were extracted to train multiple regressor models, i.e., support vector machine (SVM), multi-layer perceptron (MLP), random forest regressor (RFR), and eXtreme gradient Boosting (XGBOOST). The validation accuracies were: 0.329 for SVM, 0.414 for MLP, 0.356 for RFR and 0.429 for XGBOOST.

The proposed paper aims to establish the correlation between handcrafted features and overall survival prediction. Unlike the existing state-of-the-art methods used for survival prediction \cite{agravat2019brain},\cite{wang2019automatic}, \cite{feng2019brain}, \cite{wang20193d}, the paper uses four predictors and two feature sets to establish their correlation with overall survival prediction of High Grade Glioma (HGG) patients. Shape features and gradient boosting regressors achieve better survival prediction accuracy than state-of-the-art methods. It establishes that shape features have a strong correlation with survival prediction. The organization of the remainder of the paper is as follows: The Brain Tumor Segmentation (BraTS) dataset is described in Section 2, survival prediction methods with four predictors and two feature sets are in Section 3, Section 4 contains results and discussions and finally the conclusion of the paper is in Section 5.

\section{BraTS dataset} %\cite{bakas2017advancing,bakas2018identifying,menze2014multimodal}}
\renewcommand{\baselinestretch}{.5}
\vspace{-4mm}

Due to different standards and differences in the dataset, evaluating brain tumor segmentation methods objectively and predicting overall survival is a challenge. Nevertheless, for a comparison of different tumor segmentation and survival prediction techniques, the BraTS (brain tumor segmentation challenge) \cite{bakas2017advancing,bakas2018identifying,menze2014multimodal} has become a popular platform. Since the year 2018, there are three tasks that are included in this platform. The first task is the process of segmenting the brain tumor. The second task is predicting the overall survival (OS) and the third task is estimating the uncertainty for the predicted tumor sub-regions.  The process of tumor segmentation involves delineating the tumor into three sub-regions, namely, the whole tumor, the tumor core, and the enhancing tumor. Specificity and sensitivity metrics as well as Dice score and Hausdorff Distance are used for evaluating performance.

The overall survival prediction task classifies survival days into the following categories: long-term survivors ($>$15 months), intermediate-survivors (between 10 and 15 months), and short-survivors ($<$10 months). Samples with resection status GTR (gross total resection) are used to rate the performance of the OS prediction. An accuracy metric is used for performance evaluation, whereas mean and median square error are used for postanalysis \cite{rajput2020review}.

The BraTS 2020 training dataset includes 369 volumetric samples of high-grade glioma (HGG) and low-grade glioma (LGG) cases. It includes metadata of 236 samples such as age, survival days, and resection status for survival days prediction (Grosstotal Resection (GTR) = 119, Sub-total Resection (STR) = 10, and NA = 107). The validation dataset includes 125 sample images and metadata (age, survival days, and resection status) with 29 images having a GTR resection status.  Each subject includes four MRI scans that are preoperative (T1-weighted, T1-CE, T2-weighted, and FLAIR) and manually annotated ground truth results. The annotations of ground truth include Necrotic and Non-Enhancing tumor core NCR/NET (label-1), Edema (label-2), Active Tumor (label-4), and 0 for everything else. The dataset has been pre-processed, i.e., all the scans are co-registered to the same anatomical structure, skull stripped and resampled to an isotropic resolution of $ 1\times1\times1\hspace{0.2cm}mm^3$. The width, height, and depth of each sample are 240, 240, and 155 respectively.

\section{Survival Prediction Methodology}
\vspace{-4mm}

\begin{figure}[!h]
\begin{subfigure}{.32\textwidth}
\centering
\includegraphics[width=3cm,height=3cm]{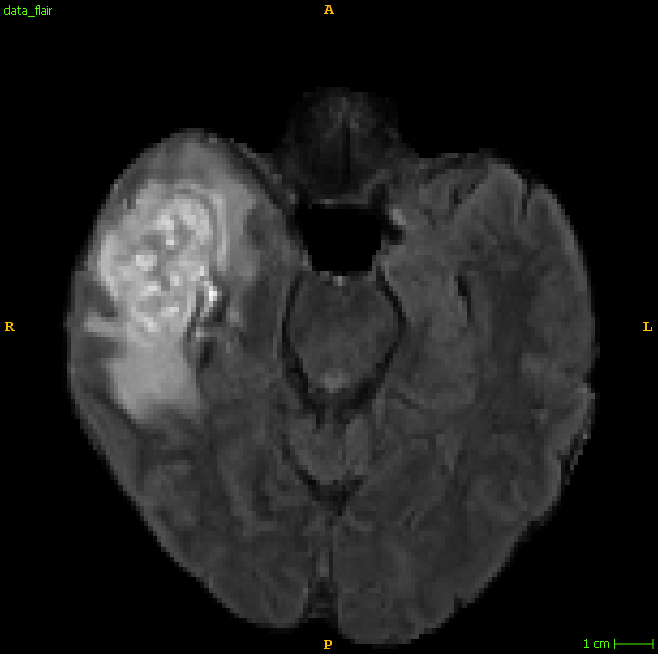}
\caption{}
\end{subfigure} \hfill
\begin{subfigure}{.32\textwidth}
\centering
\includegraphics[width=3cm,height=3cm]{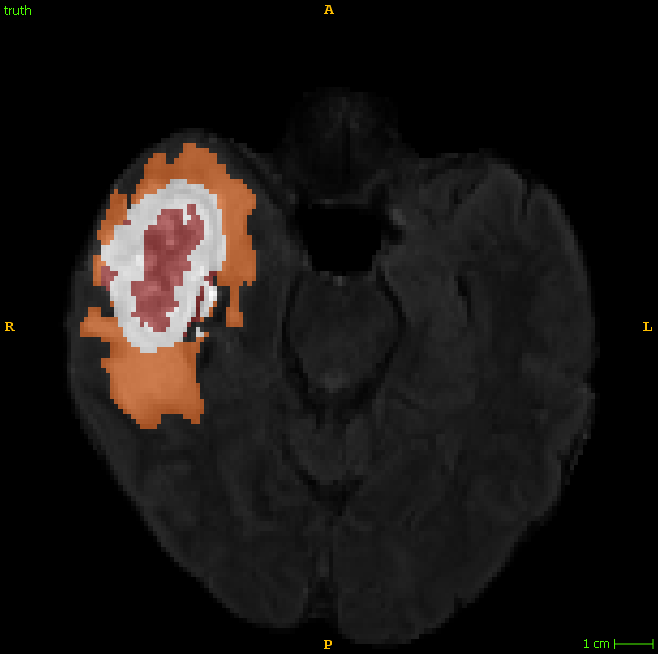}
\caption{}
\end{subfigure} \hfill
\begin{subfigure}{.32\textwidth}
\centering
\includegraphics[width=3cm,height=3cm]{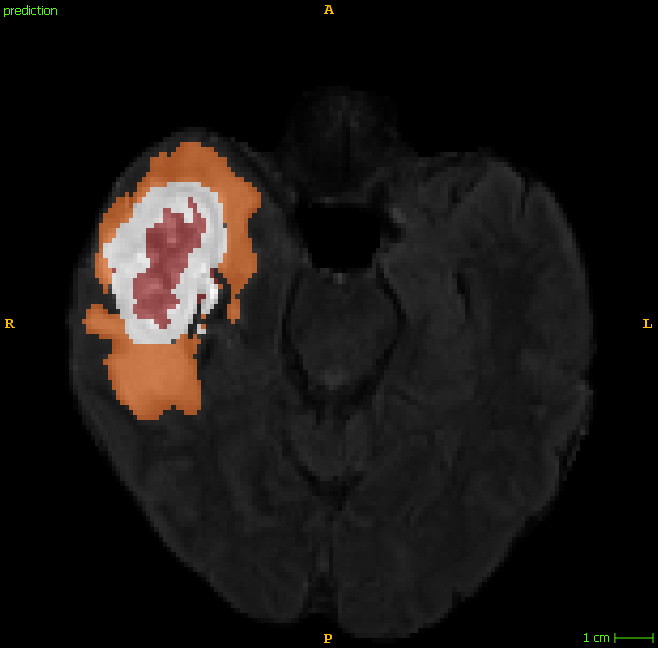}
\caption{}
\end{subfigure} \vfill

\begin{subfigure}{.32\textwidth}
\centering
\includegraphics[width=3cm,height=3cm]{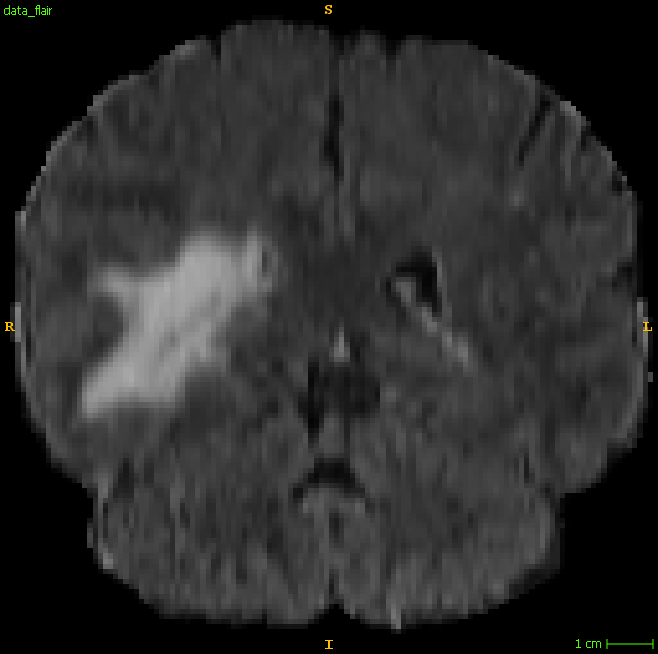}
\caption{}
\end{subfigure} \hfill
\begin{subfigure}{.32\textwidth}
\centering
\includegraphics[width=3cm,height=3cm]{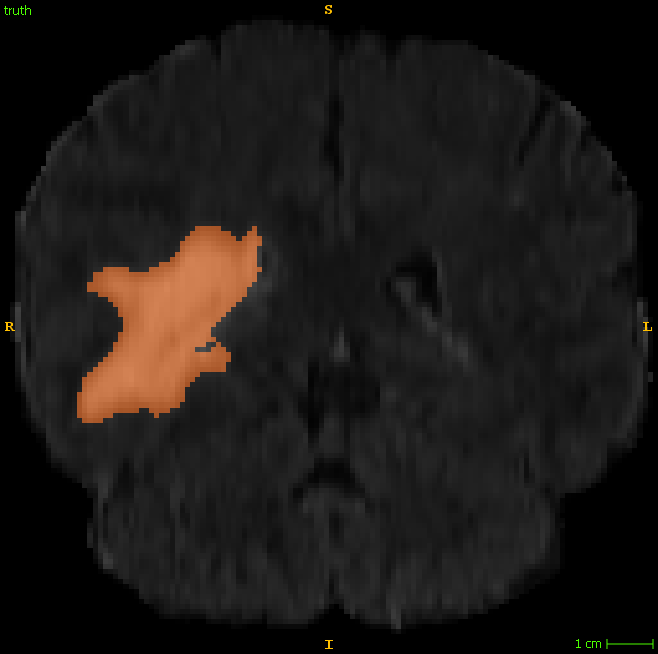}
\caption{}
\end{subfigure} \hfill
\begin{subfigure}{.32\textwidth}
\centering
\includegraphics[width=3cm,height=3cm]{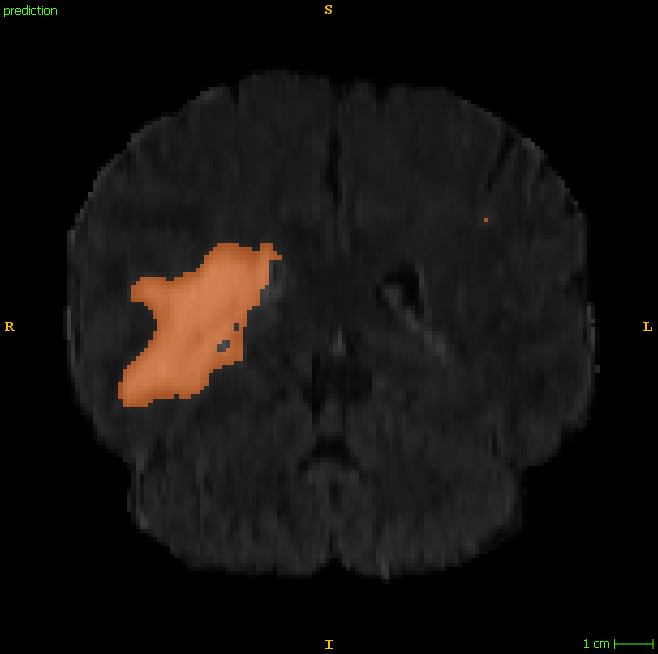}
\caption{}
\end{subfigure} \vfill

\begin{subfigure}{.32\textwidth}
\centering
\includegraphics[width=3cm,height=3cm]{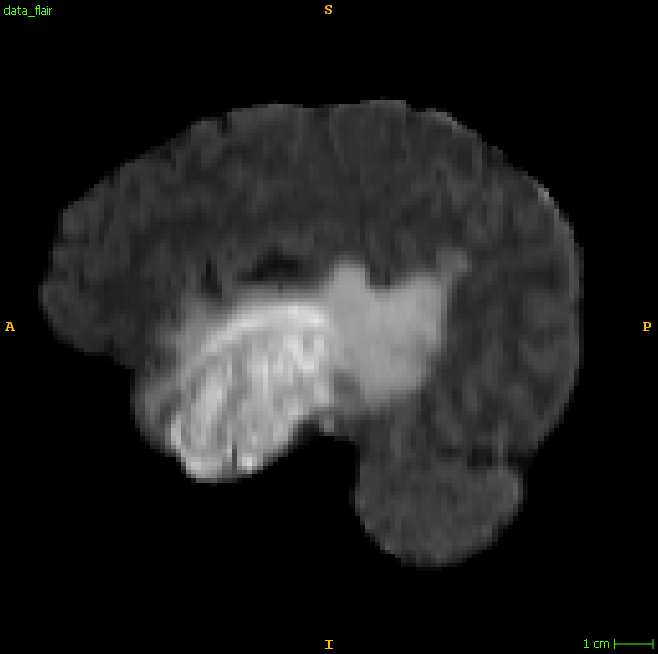}
\caption{}
\end{subfigure} \hfill
\begin{subfigure}{.32\textwidth}
\centering
\includegraphics[width=3cm,height=3cm]{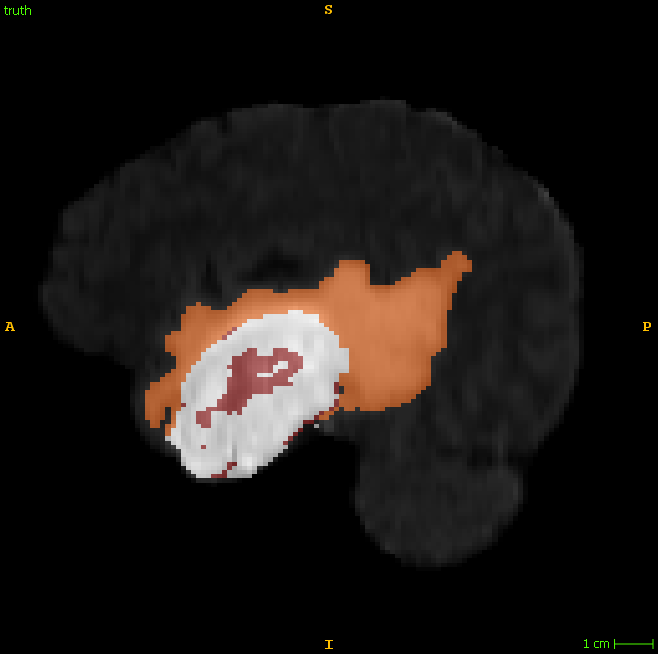}
\caption{}
\end{subfigure} \hfill
\begin{subfigure}{.32\textwidth}
\centering
\includegraphics[width=3cm,height=3cm]{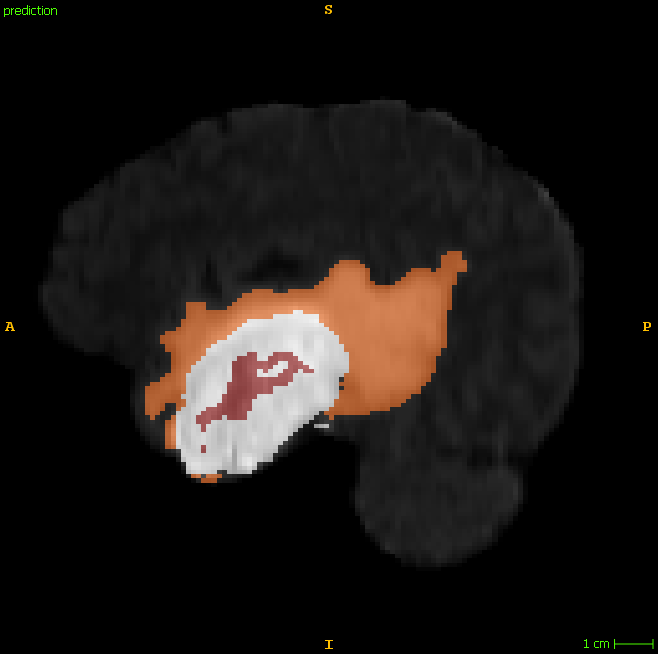}
\caption{}
\end{subfigure} 
\caption{Segmentation results of training set: (a) Axial FLAIR slice (b) Axial Ground truth (c) Axial Segmentation (d) Coronal FLAIR slice (e) Coronal Ground truth (f) Coronal Segmentation (g) Sagittal FLAIR slice (h) Sagittal Ground truth (i) Sagittal Segmentation, four color codes are: Brown for label-1(NCR/NET), white for label-4(Active Tumor), orange for label-2(Edema), black for label-0(back ground)}
\label{fig:fig3}
\end{figure}

We use the 3D U-Net model for brain tumor segmentation proposed by Isensee et al. \cite{isensee2017brain}. This is the highest ranking and simple model in BraTS 2017. Like the U-Net \cite{ronneberger2015u}, this model \cite{isensee2017brain} comprises a contracting path to extract more feature information with increasing network depth. It has an expansion path to generate a segmentation mask with precise localization information and a skip connection for better feature reconstruction at every stage of the expansion path. In our work we have used the bias field correction, normalization, clipping maximum/ minimum intensity to remove outliers, rescaled to $[0,\, 1]$ and setting non-brain pixels to 0. The model was trained on a patch size of 128×128×128, randomly generated from all the input MRI modalities. The obtained dice score on the BraTS 2020 validation dataset is 0.880(WT), 0.858(TC), 0.759(ET). The segmentation of tumor tissue of a validation sample is as shown in \ref{fig:fig3}. The figures show a visual comparison of an input flair image and a predicted image. The segmented parts are then used for survival prediction with the prognosis methods with 1) Image-based features, 2) Radiomics based features, and the following four predictors.

%\begin{figure}[!ht]
%\vspace{-5mm}
%\centering
%\begin{subfigure}[t]{0.30\textwidth}
%    \makebox[0pt][r]{\makebox[10pt]{\raisebox{40pt}{\rotatebox[origin=c]{90}{Flair %image}}}}%
%    \includegraphics[width=\textwidth]
%    {templates/fig/val-axial-flair-103.png}
%    
%    \makebox[0pt][r]{\makebox[10pt]{\raisebox{40pt}{\rotatebox[origin=c]{90}{Predicted %Image}}}}%
%    \includegraphics[width=\textwidth]
%    {templates/fig/val-predict-axial-103.png}
%    \caption{Axial}
%\end{subfigure}
%\hspace{1em}
%\begin{subfigure}[t]{0.30\textwidth}
%    \includegraphics[width=\textwidth]  
%    {templates/fig/val-coronal164.png}
%    
%    \includegraphics[width=\textwidth]
%    {templates/fig/val-predict-coronal164.png}
%    \caption{Coronal}
%\end{subfigure}
%\hspace{1em}
%\begin{subfigure}[t]{0.30\textwidth}
%    \includegraphics[width=\textwidth]  
%    {templates/fig/val_sagittal_105.png}
%   
%    \includegraphics[width=\textwidth]
%    {templates/fig/val-predict-sagittal105.png}
%    \caption{Sagittal}

%\end{subfigure}

%\caption{Tumor Segmentation from the validation dataset and the corresponding  predicted image. For predicted image, four color codes are: brown for label-1(NCR/NET), white for label-4(Active Tumor), orange for label-2(Edema), black for label-0(back ground).}

%\label{fig:fig3}
%\vspace{-9mm}
%\end{figure}
%=====================

\subsection{Predictors and Parameter Tuning}

We have used four predictors and parameter tuning. These are (1) Artificial Neural Network (ANN) \cite{wang20193d,islam2019brain}, (2) Linear Regressor (LR) \cite{wang2019automatic,feng2019brain}, (3) Gradient Boosting Regressor (GBR) \cite{islam2019brain}, and (4) Random Forest Regressor (RFR) \cite{agravat2019brain,isensee2017brain,islam2019brain}. All these predictors were used by the top performing models in all recent BraTS challenges. These predictors deal with a small dataset and overfitting problems. The image-based prognosis method uses only seven features making it less vulnerable to overfitting. We retain default parameters for ANN and LR, while parameters for GBR and RFR are hyper-tuned using a grid search. We tuned the number of estimators, depth of the tree, sample split, and learning rate parameters for the GBR. In the case of the RFR, the number of estimators and the depth of the tree were hyper tuned. The predictors with radiomics features were also tuned.

For radiomics features it turns out that an ANN with five hidden layers was better compared to 2 or 3 hidden layers. Further, we tuned epochs, learning rate, number of neurons, and an optimizer for ANN. In the LR model, a search was also performed for the penalty term, the number of iterations, and up-grading of feature parameters using LASSO and a ridge regressor. We tuned the number of estimators, maximum depth, and learning rate for the GBR. In the RFR model, we tuned the number of estimators, maximum depth of the tree, minimum sample split, minimum samples in a leaf node, and maximum features parameters. Since the random forest and gradient boosting regressor work on ensemble-based learning, they are robust, efficient, and less prone to overfitting.

\subsection{Prognosis using Features}

\subsubsection{Image-based features \cite{feng2019brain,wang20193d}}

 Shape features extracted from the segmentation were used in the OS prediction. These features were volume of the WT, TC, and ET, surface area of the WT, TC, and ET, age. Since the tumor size was the decisive predicting factor for various cancer types, we extracted the volume and surface area of the WT, TC, and ET. The features were extracted from the segmentation maps and input images without any library dependency. Training with fewer features has the advantage that it limits the dimensions of feature space. Hence, the model did not overfit. However, we found saturation in the performance due to high bias in the model.

\subsubsection{Radiomics based features \cite{van2017computational}}

Radiomics based feature extraction is widely used for disease diagnosis, classification, and survival prediction like lung cancer \cite{he2018biomarker}, breast cancer \cite{liu2019preoperative}, and Alzheimer’s disease \cite{li2019radiomics}. Along with the size of the tumor, exploring the correlation of the other features with survival prediction is crucial to increase the performance of the predictor models. Radiomics features addresses this problem. It allows extracting various statistical, shape, intensity, and texture features from radiographic scans. Also, radiomics allow extracting features from many imaging techniques.

Using the package PyRadiomics \cite{van2017computational}: the following 107 features were extracted:
\begin{enumerate}
    \item Shape features: Elongation, flatness, axis lengths, maximum diameter, mesh volume, sphericity, surface area, and surface volume ratio.
    \item Gray level features: Gray-level size zone (GLSZ), Gray-level co-occurrence matrix (GLCM), Gray-level run-length matrix (GLRLM), Gray-level dependence Matrix (GLDM), and neighbouring gray-tone difference matrix (NGTDM).
    \item First-order statistical features: Energy, entropy, minimum intensity value, maximum intensity value, mean, median, Interquartile range, percentiles, absolute deviation, skewness, variance, kurtosis, and uniformity.
\end{enumerate}

Radiomics features are typically multi-collinear and redundant \cite{weninger2019robustness}; hence the correlation between these features needs to be validated for specific real-world problems. We performed feature selection through recursive feature elimination (RFE) \cite{pedregosa2011scikit} to remove weaker features and avoid the curse of dimensionality. RFE is an example of backward feature elimination. With the given number of estimators, it selects principal features recursively from the feature set. It refits the model until the desired number of selected features is eventually reached. Out of 107 features, we selected 20 best ranking features.

In summary, the four predictors: ANN, RFR, LR, and GBR, are applied to: i) the seven image-based features, ii) 107 radiomics features, iii) 20 principal radiomics features, and iv) only shape radiomics features. Literature \cite{agravat2019brain,isensee2017brain} also suggests dominance of shape features so we also used all predictors with only shape features for survival prediction. We trained the models with all the resection status (i.e., GTR, STR, and NA) given with the dataset to increase the database size and reduce overfitting.

\section{Results and Discussions}
\vspace{-2mm}

Image-based feature prediction is derived from the BraTS 2019 dataset, and the BraTS 2020 dataset was used for radiomics based feature extraction. The results are shown in Tables 1 to 4. We have not participated in the BraTS 2020 challenge and do not have access to the test dataset. Therefore, results are derived on the training and validation datasets.

\subsection{Image-based feature prediction}
\vspace{-1mm}

We observe that the ensemble-based models, i.e., GBR and RFR, show a better performance on the training and validation dataset. Their consistency in the training and validation accuracy suggests that the model does not overfit.

% \usepackage{multirow}
\begin{table}[]
\vspace{-5mm}
\caption{OS Performance comparison using image-based feature on training and validation BraTS-2019 dataset. MSE, medianSE, stdSE, and SpearmanR denote the mean square error, median square error, standard deviation squared error, and Spearman's ranking coefficient.}
\label{tab:table1}
\centering
\begin{tabular}{ccccccc}
\toprule
Dataset & Regressor & Accuracy & MSE & medianSE & stdSE & SpearmanR \\
  \midrule
\multirow{4}{*}{Training}   & ANN & 0.51 & 86148.10 & 21316 & 181346 & 0.48 \\
                            & LR  & 0.49 & 87724.00 & 20736 & 183685 & 0.47  \\
                            & GBR & 0.52 & 63234.40 & 16900 & 126534 & 0.61 \\
                            & RFR & 0.52 & 63234.40 & 16900 & 126534 & 0.61 \\
\midrule
\multirow{4}{*}{Validation} & ANN & 0.45 & 098312.70 & 39204 & 141392 & 0.24 \\
                            & LR  & 0.52 & 100509.00  & 38809 & 141263 & 0.29 \\
                            & GBR & 0.52 & 102999.00  & 36481 & 152694 & 0.27 \\
                            & RFR & 0.52 & 102999.00  & 36481 & 152694 & 0.27 \\
\bottomrule
\end{tabular}
\vspace{-2mm}
\end{table}

\subsection{Radiomics feature-based prediction}

As mentioned, we extracted 107 radiomic features from the segmentation results of the BraTS 2020 images and fed them as input to four regressor models; ANN, LR, GBR, and RFR. It was observed that RFR gave the best results, and they are shown in Table \ref{tab:table2}. The other regressors performed poorly compared to RFR, and even the fine-tuning of the parameters did not improve the performance. The possible reasons are the redundant nature of radiomics \cite{weninger2019robustness}, over complexity due to too many features and fewer training samples. Radiomics features are shallow and low-order image features, and unable to fully describe distinct image characteristics \cite{lao2017deep}. Also, when the number of observations is less for large extracted features, survival prediction is an ill-posed problem \cite{weninger2019robustness}.

\begin{table}[]
\vspace{-2mm}
\centering
\caption{OS performance evaluation using 107 radiomics features and Random Forest Regressor. }
\label{tab:table2}
\begin{tabular}{p{2cm}p{2cm}p{2cm}p{2cm}p{2cm}p{2cm}}
\toprule
Dataset    & Accuracy & MSE      & medianSE & stdSE     & SpearmanR \\
\midrule
Training   & 0.479    & 079176.96 & 20702.21 & 169474.53 & 0.684     \\
Validation & 0.379    & 115424.30 & 28779.30  & 214028.11 & 0.138    \\
\bottomrule
\end{tabular}
\vspace{-5mm}
\end{table}

It can be observed from Table \ref{tab:table2} that the large feature set is unable to yield state-of-the-art accuracy results. Therefore, we reduced the feature set by applying recursive feature elimination to find the 20 most dominant features. Dominant features obtained using RFE are: age, amount of edema, elongation, maximum 2D diameter slice, sphericity, surface-volume ratio, minimum and maximum intensity, interquartile range, skewness, kurtosis, root mean absolute deviation, cluster prominence, cluster shade, inverse variance, coarseness, and dependence variance. We then applied four regressors on the dominant feature set, and performance has been noted in Table \ref{tab:table3}.

\begin{table}[]
\vspace{-7.7mm}
\centering
\caption{OS performance comparison on 20 principal radiomics features.}
\label{tab:table3}
\begin{tabular}{p{1.5cm}p{1.5cm}p{1.5cm}p{1.5cm}p{1.5cm}p{1.5cm}p{1.5cm}}
\toprule
Dataset  & Regressor  Models & Accuracy & MSE      & medianSE   & stdSE     & SpearmanR \\
\midrule
\multirow{4}{*}{Training}   & ANN   & 0.393 & 8.90E+12 & 2.46E+12  & 3.36E+13  &  0.125     \\
                            & LR    & 0.462    & 96853.55 & 33279.52  & 190733.00  & 0.417     \\
                            & GBR   & 0.923    & 17213.25 & 00000.00   & 074717.13  & 0.938     \\
                            & RFR   & 0.744    & 31829.75 & 06077.32  & 075572.44  & 0.810  \\
\midrule
\multirow{4}{*}{Validation} & ANN  & 0.448    & 2.20E+20 & 3.46E+12  & 8.03E+20  & 0.290 \\
                            & LR & 0.483    & 2.73E+08 & 056167.55  & 9.86E+08  & 0.456     \\
                            & GBR & 0.414    & 255096.40 & 101995.06 & 420861.25 & 0.025     \\
                            & RFR & 0.448    & 098369.46 & 035521.48  & 126218.18 & 0.126     \\
\bottomrule
\end{tabular}
\vspace{-5mm}
\end{table}

We observe that the linear regressor with regularisation outperforms all other regression models with the highest accuracy on the validation dataset. LR also provides similar accuracy for the training and validation datasets. The Spearman-R is also highest for LR. In contrast, RFR achieves the lowest mean square error (MSE) on the validation dataset.

\subsubsection{Radiomic shape features based prediction}

Reviewing the correlation between radiomics features and survival prediction, we found that radiomic shape features play a crucial role in survival prediction \cite{agravat2019brain,isensee2017brain}. Shape features show significant statistical differences across ROIs \cite{chaddad2016quantitative}. Hence, shape features can capture tumor features related to genetic anomalies and profoundly impact survival prediction. We formulate the hypothesis that \emph{shape features profoundly impact survival prediction}. In order to validate the hypothesis, we trained predictor models with the following shape features: the amount of necrotic, edema, enhancing tumor, the extent of the tumor, coordinates of tumor, elongation, flatness, axis lengths, 2D diameter row, 2D diameter column, 2 D diameter slice, maximum 3D diameter, mesh volume, sphericity, surface area, surface volume ratio, centroid of necrosis and age information. The performance of each predictor model has been noted in Table \ref{tab:table4}. 

\begin{table}
\vspace{-5mm}
\centering
\caption{OS performance comparison on BraTS-2020 dataset using radiomics shape features set.}
\label{tab:table4}
\begin{tabular}{p{1.5cm}p{1.5cm}p{1.5cm}p{1.5cm}p{1.5cm}p{1.5cm}p{1.5cm}}
\toprule
Dataset & Predictor Models & Accuracy & MSE      & medianSE & stdSE     & SpearmanR \\
\midrule
\multirow{4}{*}{Training}   & ANN  & 0.400      & 4.41E+11 & 7.15E+10 & 7.97E+11  & 0.149     \\
                            & LR & 0.470     & 89890.41 & 35160.09 & 162137.20  & 0.461     \\
                            & GBR  & 0.915    & 31068.75 & 00000.00        & 150724.63 & 0.849     \\
                            & RFR  & 0.615    & 62930.78 & 18562.88 & 130788.18 & 0.759     \\
\midrule
\multirow{4}{*}{Validation} & ANN  & 0.448    & 4.73E+11 & 2.14E+11 & 5.97E+11  & 0.149     \\
                            & LR   & 0.414    & 087228.24 & 47820.00    & 111960.30  & 0.215     \\
                            & GBR  & 0.621    & 141065.30 & 23528.48 & 236728.70  & 0.338     \\
                            & RFR  & 0.448    & 109746.60 & 34689.29 & 200725.98 & 0.116   \\ 
\bottomrule
\end{tabular}
\vspace{-5mm}
\end{table}
\raggedbottom

We observe that GBR and RFR have better performance. Specifically, the gradient boosting regressor outperforms all other regression models. In contrast, LR with regularization achieves the lowest mean square error (MSE) on the validation dataset.

\subsection{Discussions}

It has been observed that classical machine learning techniques performed better than the deep learning neural network-based models for survival prediction. Radiomics based approaches are well suited for survival prediction. Traditional regression algorithms have better interpretability than deep learning-based algorithms, they have fewer learnable parameters than CNN, and perform better with smaller sample data. A large sample dataset for training is crucial for direct regression from image modalities using CNN.

The predictors trained on the 107 radiomics features underperformed. The predictors modelled on the 20 principal features improved the performance. Further, to alleviate performance, we experimented and trained predictors on shape features and found a strong correlation with survival prediction. Shape features trained on the consensus model obtained state-of-the-art survival prediction accuracy. It was observed that the gradient boosting regressor model performed better than other classical algorithms because of: additive model, and with each tree built, the model becomes more expressive based on the ensemble learning model. The proposed GBR model is compared with the survival prediction challenge winners of BraTS 2020 and prediction accuracy for the state-of-the-art methods was obtained from the unranked leader board\footnote{https://www.cbica.upenn.edu/BraTS20/lboardValidation.html}. TA performance comparison of the GBR model with top-ranking models has been noted in Table \ref{tab:table5}. It can be observed that shape-based features with the gradient boosting regressor outperform the best-ranking methods over the validation dataset.

\begin{table}[]
\vspace{-6mm}
\centering
\caption{OS performance comparison with top-ranking models on the BraTS-2020 validation dataset.}
\label{tab:table5}
\begin{tabular}{p{2.5cm}p{1.5cm}p{2cm}p{2cm}p{2cm}p{2cm}}
\toprule
\textbf{Team name}                    & \textbf{Accuracy} & \textbf{MSE}      & \textbf{medianSE} & \textbf{stdSE}    & \textbf{SpearmanR} \\
\midrule
SCAN         & 0.414 & 098704.65  & 36100.00    & 152175.57 & 0.253 \\
Redneucon    & 0.517 & 122515.76 & 70305.26 & 157673.99 & 0.134 \\
VLB          & 0.379 & 093859.54  & 67348.26 & 102092.41 & 0.280  \\
COMSATS-MIDL & 0.483 & 105079.42 & 37004.93 & 146375.99 & 0.134 \\
\textbf{Proposed} & \textbf{0.621}    & 141065.30 & \textbf{23528.40} & 236728.70 & \textbf{0.338}    \\
\bottomrule
\end{tabular}
\vspace{-7mm}
\end{table}

\section{Conclusion}

Predicting oncological outcomes is always very tricky due to multiple challenges from clinical and engineering perspectives. In this work, we have evaluated two feature sets over four predictors. We proposed the image-based and the radiomic based prognosis approaches for survival prediction. The image-based prognosis models performed well, but the performance saturates beyond a certain point because of fewer features, and models could not learn complexity. Similar observations are also made for the 107 radiomics features / 20 principal features and the regressor combination. All above the combinations exhibited correlation with survival prediction. However, we recommend that shape based features with the gradient boosting regressor is the best combination for survival prediction. Comparing models, it was found that ensemble-based learning models became more useful for survival prediction because of their robustness. Whereas ANN converges speedily compared to classical models but due to lack of ample training samples, it overfits easily. With the availability of a large dataset and more clinical non-imaging information such as gender and treatment, survival prediction can be robust. It can further be applied to clinical practice.

\bibliographystyle{splncs04}
\bibliography{bibref}
\end{document}